\documentclass[11pt,a4paper]{article}

\usepackage{booktabs}
\usepackage[utf8]{inputenc}
\usepackage[T1]{fontenc}
\usepackage[margin=1in]{geometry}
\usepackage{amsmath, amssymb, amsfonts, bm}
\usepackage{graphicx}
\usepackage{algpseudocode}
\usepackage{algorithm}
\usepackage{array}
\usepackage{multirow}
\usepackage{subcaption}
\usepackage{float}
\usepackage{color, soul}

\usepackage[numbers]{natbib} % Use [numbers] as requested

\usepackage[colorlinks=true, linkcolor=blue, citecolor=blue, urlcolor=blue]{hyperref}

\newcommand{\copyrightnotice}{%
 \thanks{This is a Gold Open Access article made available under the CC-BY license. 
 Published in \textit{Sustainable Energy, Grids and Networks}. 
 DOI: \href{https://doi.org/10.1016/j.segan.2025.102009}{10.1016/j.segan.2025.102009}}
}

\title{\textbf{Profit Maximization for Electric Vehicle Charging Stations Using Multiagent Reinforcement Learning}\copyrightnotice}

\author{
    Kun-Yan Jiang$^{1}$, 
    W.-Y. Chiu$^{2,}$\thanks{Corresponding author: weiyu.chiu@deakin.edu.au}, 
    Yuan-Po Tsai$^{1}$ \\
    \small $^{1}$Department of Electrical Engineering, National Tsing Hua University, Hsinchu, Taiwan \\
    \small $^{2}$School of Information Technology, Deakin University, Melbourne, Victoria, Australia 
}
\date{}

\begin{document}

\maketitle

\begin{abstract}
Electric vehicles (EVs) are increasingly integrated into power grids, offering economic and environmental benefits but introducing challenges due to uncoordinated charging. This study addresses the profit maximization problem for multiple EV charging stations (EVCSs) equipped with energy storage systems (ESS) and renewable energy sources (RES), with the capability for energy trading. We propose a Double Hypernetwork QMIX-based multi-agent reinforcement learning (MARL) framework to optimize cooperative energy management under uncertainty in EV demand, renewable generation, and real-time electricity prices. The framework mitigates overestimation bias in value estimation, enables distributed decision-making, and incorporates an internal energy trading mechanism. Numerical experiments using real-world data demonstrate that, compared to standard QMIX, the proposed method achieves approximately 5.3 \% and 12.7 \% higher total profit for the two regions, respectively, highlighting its economic and operational efficiency. Additionally, the approach maintains robust performance under varying levels of EV demand uncertainty and renewable energy fluctuations.
\end{abstract}


% --- Main Content ---
\newpage

\section{Introduction}
\label{sec1}

Electric vehicles (EVs) have gained significant attention recently due to their potential to offer substantial economic and environmental benefits \citep{21Li}. These advantages include the alleviation of environmental concerns associated with fossil fuels and the mitigation of grid instability stemming from the integration of renewable energy generation \citep{19Yongmin}. However, the uncoordinated charging behavior of a large number of EVs can introduce considerable negative impacts on power grids. These impacts range from increased system losses and voltage drops to phase imbalances and broader stability issues \citep{18Yu, 20Amro, 16Junjie}.

To address these challenges, extensive research has been conducted on the management of individual EV charging stations (EVCSs). The primary focus of this body of work has been on optimizing charging and discharging events to minimize costs or maximize profits for a single EVCS \citep{18Abdorreza}. In typical models, an EVCS communicates electricity prices to EV users, who then formulate their charging plans and share them with the station \citep{18Xiaoxuan}. These models can also accommodate the preferences of EV users and urgent charging scenarios \citep{18Jean}. In practice, however, EVs are mobile and travel between different regions, connecting to various EVCSs for recharging \citep{19Mustafa}. This mobility necessitates a coordinated approach to manage charging across multiple EVCSs on a larger scale.

In a coordinated framework, multiple EVCSs, often equipped with energy storage systems (ESS) and renewable energy sources (RES) \citep{19Pilar, 18Kalpesh}, can operate with greater flexibility. A central coordinator can facilitate this by coordinating the activities of EVCSs and EVs. An effective coordinating strategy for charging and discharging events can significantly reduce the operating costs for both the grid and the EVCSs themselves \citep{18Mushfiqur}. Consequently, cooperative schemes for EVCSs have been explored, involving the trading of energy and the sharing of information related to EV energy demand, photovoltaic (PV) generation, and ESS battery capacity \citep{16James, 17Joy}.

To manage the uncertainty inherent in such dynamic systems, learning-based methods, particularly reinforcement learning (RL), have emerged as a powerful tool. RL allows for the presence of unknown system parameters while still achieving a designated goal \citep{shakya2023reinforcement}. The uncertainty within a system can be treated as an unknown distribution from which RL agents can learn. By interacting with a stochastic environment, RL agents can effectively tackle system uncertainty and maximize an expected cumulative reward over a learning process \citep{shahab2024designing}. For the management of a single EVCS, single-agent RL has been successfully employed. For instance, \cite{18Shuoyao} investigated an optimal pricing and scheduling strategy to maximize the profit of an EVCS, while \cite{18Sunyong} utilized RL-based energy management to reduce the operating energy cost of a smart building by considering EV charging activities.

However, a single geographical region typically hosts multiple EVCSs, leading to multi-agent challenges arising from their mutual interactions \citep{huang2024multi}. In such a multi-agent scenario, optimizing energy scheduling at each EVCS becomes crucial to ensure the efficient use of resources, accommodate fluctuating charging demands, and adapt to dynamic factors like renewable energy availability and real-time electricity prices. Multi-agent reinforcement learning (MARL) provides a robust framework for this by combining the architecture of a multi-agent system with the optimization capabilities of RL. Within the MARL domain, researchers have explored various distributed Q-learning methodologies, such as Independent Q-Learning (IQL) and the centralized training with decentralized execution (CTDE) paradigm \citep{bai2023towards}.

Recent research has increasingly focused on applying MARL techniques  \citep{2022Lai} to optimize the profitability of EVCSs. In terms of coordination among multiple agents in EVCSs, researchers have adopted two primary approaches: cooperative and non-cooperative. For instance, \cite{qian2021multi} proposed a dynamic pricing strategy for multiple EVCSs to maximize the profit of individual stations by considering the EV load profile in a non-cooperative game setting. On the cooperative front, \cite{adetunji2023two} utilized a centralized architecture to manage a non-stationary environment for multiple EVCSs. \cite{20Valeh} proposed an online pricing strategy to control EV charging demands and bolster utility stability while increasing station revenue. Furthermore, \cite{20Felipe} presented a hybrid approach, combining cooperative and non-cooperative strategies to minimize energy costs and prevent transformer overloads.

Although significant progress has been made in the application of MARL for EVCS energy management, several gaps remain. Many studies focus either on cooperative or non-cooperative settings, but rarely incorporate mechanisms that ensure  robust value estimation under  uncertain environments. Conventional MARL algorithms such as IQL or basic QMIX often suffer from overestimation bias in value functions, which can degrade policy quality and lead to unstable coordination outcomes. Moreover, current works largely overlook the integration of energy trading mechanisms among EVCSs, which could enhance flexibility and profitability while reducing reliance on the utility grid. These gaps underscore the need for a robust MARL framework that explicitly addresses overestimation bias, supports cooperative decision-making, and incorporates energy trading as a core component of system optimization.

This article investigates a profit maximization problem for the coordination of multiple EVCSs, each equipped with an ESS and RES, with the ability to trade energy amongst themselves. To tackle the uncertainty introduced by real-time pricing, renewable generation, and EV energy demand, we propose a double hypernetwork QMIX-based MARL framework tailored for profit maximization in cooperative EVCS energy management.

The main contributions of this article are summarized as follows:
\begin{itemize}
    \item \textit{Cooperative and Distributed Architecture for EVCSs:} We propose a distributed approach for coordinating multiple EVCSs, which offers enhanced scalability and resilience by reducing a single point of failure, a departure from centralized methods. While a coordinator facilitates the exchange of aggregate information, it does not dictate local decision-making, ensuring that each EVCS can operate autonomously if needed.
    \item \textit{MARL Architecture 
    and Energy Trading Mechanism:} 
    The proposed design called double hypernetwork QMIX  improves upon standard QMIX by mitigating overestimation bias in value estimation. We also integrate an internal energy trading layer that enables EVCSs to balance surplus and deficit energy before interacting with the utility grid, thereby reducing costs and improving overall profitability.
    \item \textit{Efficient Information Exchange:} In contrast to methods requiring agents to exchange detailed individual state and reward data, our approach relies on the dissemination of aggregate information, such as total EV energy demand and utility price signals. This reduction in data exchange lowers computational overhead and memory usage.
    \item \textit{Integration of Renewable Energy and Validation:} Our scheme integrates renewable energy sources, a critical component for sustainability. We validate our method using real-world data on utility prices and PV generation, demonstrating its superior performance in profit maximization and practical applicability.
\end{itemize}

The remainder of this article is organized as follows. Section~\ref{sec_model} provides the related work regarding MARL in EVCS energy management. Section~\ref{sec_model} describes our system models, including EV energy demand, models at EVCSs, energy trading models between EVCSs, and the profit model.  Section~\ref{sec_MARL2}
presents the proposed  MARL algorithm for profit maximization.
 Numerical results are given in Section~\ref{sec_sim}, including comparisons with NLP methods and conventional RL algorithms.
 Finally, Section~\ref{sec_con} concludes this article.

\section{System Models at EVCSs}\label{sec_model}

\begin{figure}[tp]
   \centering
   \begin{subfigure}[t]{0.8\textwidth}
       \centering
       \includegraphics[width=8cm]{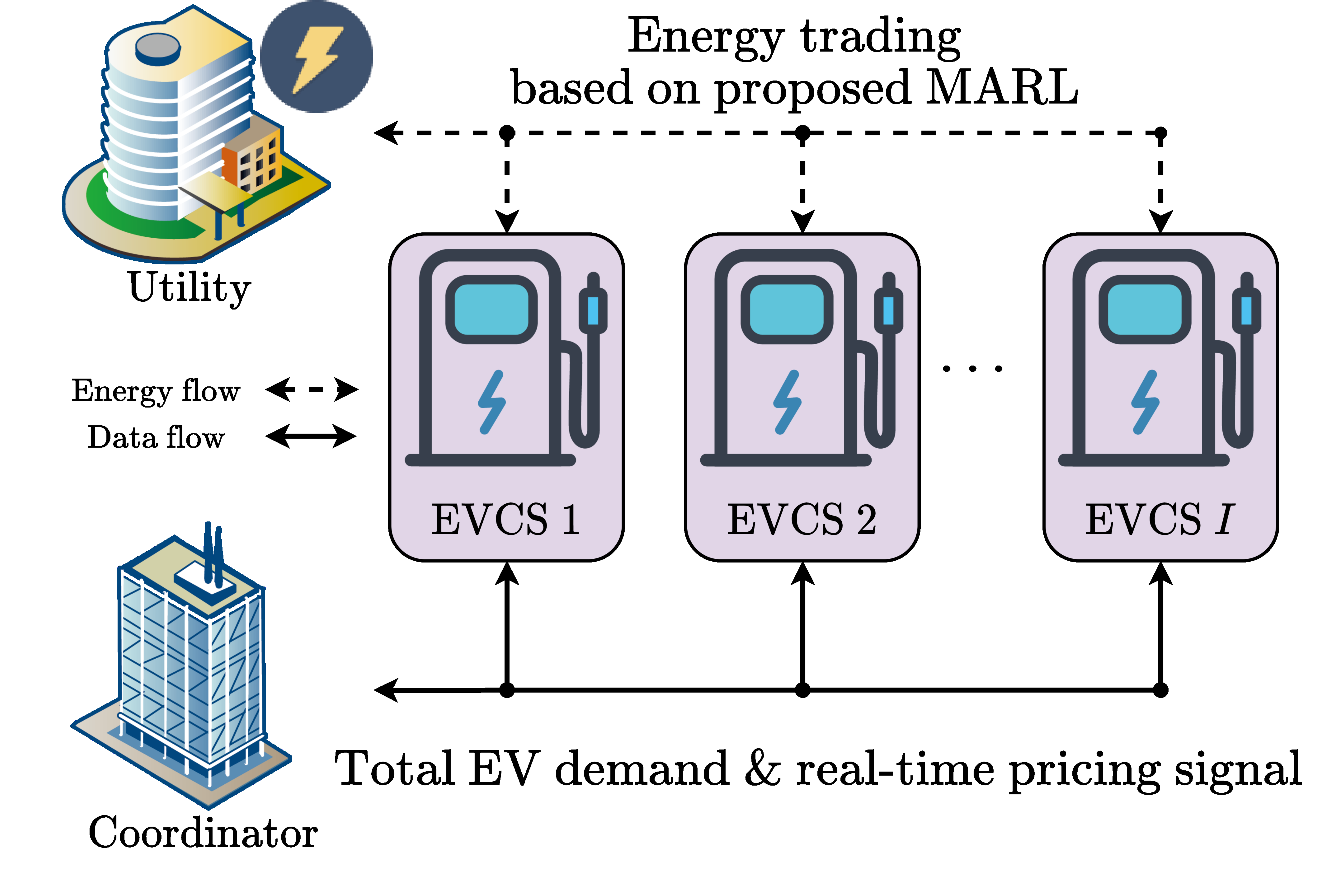}
       \caption{System architecture for multiple EVCSs.}
   \end{subfigure}
   %\vspace{0.5cm} % 增加垂直間距
   \begin{subfigure}[t]{0.8\textwidth}
       \centering
       \includegraphics[width=8cm]{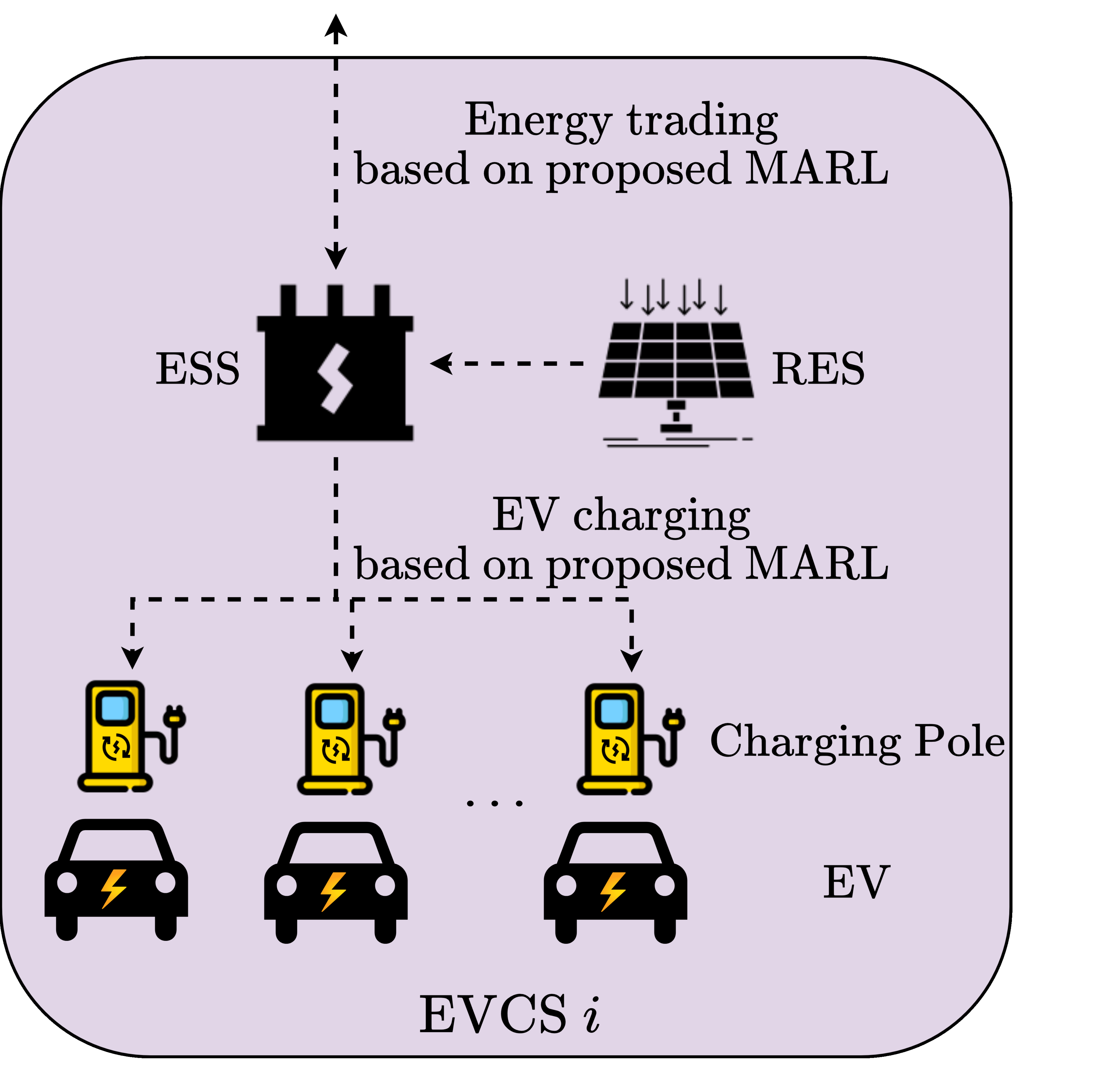}
       \caption{Components in one EVCS.}
   \end{subfigure}
   \caption{Overview of EVCS systems and components.}
   \label{fig_EVCS}
\end{figure}

Fig.~\ref{fig_EVCS}  presents the system architecture for multiple EVCSs in a microgrid, including a utility, a coordinator, EVCSs, and a group of EV fleets. The coordinator, EVCSs, and EVs can communicate with each other.
The coordinator exchanges information between EVCSs. Each EVCS has several charging poles and is equipped with an ESS and RES.
EVCSs provide charging power to meet EV energy demands and share information with each other through the coordinator.
To maximize the total profit of EVCSs, the EVCSs can purchase electricity and resell it to EVs or the utility.
This section describes the model of EV energy demand, models at EVCSs, energy trading model, and profit model in detail.

\subsection{EV Energy Demand}
EV energy demand can be classified as urgent demand and regular demand, reflecting the preference of EV users.
Urgent demand occurs when users need to quickly charge their EVs.
The amount of urgent demand and regular demand at the $i$th EVCS are denoted as $D_{\text{ev},i}^{\text{U}}(t)$ and $D_{\text{ev},i}^{\text{R}}(t)$, respectively.
Suppose that there are $I$ EVCSs;
 the total EV energy demand at the $i$th EVCS can be expressed as \citep{18Linda}
\begin{equation}\label{EV_demand}
D_{\text{ev},i}^{\text{ALL}}(t) = D_{\text{ev},i}^{\text{U}}(t)+D_{\text{ev},i}^{\text{R}}(t), i = 1, 2, ..., I.
\end{equation}
The energy supply from  the $i$th EVCS to EVs is denoted as $E_{\text{ev},i}(t)$, which  must satisfy urgent demand:
\begin{equation}\label{meet_demand bound}
\begin{split}
D_{\text{ev},i}^{\text{U}}(t) \leq E_{\text{ev},i}(t) \leq D_{\text{ev},i}^{\text{ALL}}(t).
\end{split}
\end{equation}
EVCSs can determine how much $D_{\text{ev},i}^{\text{R}}(t)$ to meet on the basis of their battery capacity, denoted as $E_{\text{b},i}(t)$.
If $E_{\text{ev},i}(t) < D_{\text{ev},i}^{\text{ALL}}(t)$, then
the unmet demand $D_{\text{ev},i}^{\text{ALL}}(t) - E_{\text{ev},i}(t) $ will become urgent demand in the next time slot $t+1$.

\subsection{Models at EVCSs}

Define the internal demand of the $i$th EVCS as
\begin{equation}\label{delta_Eb}
E_{\text{internal},i}(t)=E_{\text{r},i}(t)-E_{\text{ev},i}(t)
\end{equation}
where $E_{\text{r},i}(t)$ represents the renewable energy.
If $E_{\text{internal},i}(t)>0$, then renewable energy is sufficient to meet demand $E_{\text{ev},i}(t)$ and the ESS can store the surplus renewable energy; otherwise, renewable energy is insufficient to meet demand $E_{\text{ev},i}(t)$ and the EVCS must discharge its ESS for compensation.

The state of charge (SOC) of the ESS at the $i$th EVCS is defined as:
\begin{equation}\label{SOC}
SOC_{i}(t)=\frac{E_{\text{b},i}(t)}{E_{\text{b},\max}}, t = 1, 2, ..., T
\end{equation}
where
$E_{\text{b},i}(t)$ and $E_{\text{b},\max}$ are the battery current energy level and maximum battery capacity, respectively.
The SOC satisfies~\citep{19Chang}:
\begin{equation}\label{SOC_bound}
\begin{split}
SOC_{\min} \leq SOC_{i}(t) \leq SOC_{\max}
\end{split}
\end{equation}
where $SOC_{\text{min}}$ and $SOC_{\text{max}}$ represent the minimum and maximum SOC of the ESS, respectively.

The energy storage dynamics can be described as \citep{18Kalpesh}
\begin{equation}\label{Eb_dynamic}
\begin{split}
E_{\text{b},i}(t+1)=\beta E_{\text{b},i}(t)+E_{\text{cs},i}(t)+E_{\text{internal},i}(t)
\end{split}
\end{equation}
where $\beta \in (0, 1]$ is a coefficient associated with the power leakage,
and $E_{\text{cs},i}(t)$ represents
charging or discharging control.
 $E_{\text{cs},i}(t)>0 $ implies a charging activity with energy bought from the utility or other EVCSs;
 $E_{\text{cs},i}(t)<0$ implies a discharging activity with energy sold back to the utility or other EVCSs.
$E_{\text{cs},i}(t)$ must satisfy the following constraints:
\begin{equation}\label{action bound}
E_{\text{cs},i}^{\text{lower}}(t) \leq E_{\text{cs},i}(t) \leq E_{\text{cs},i}^{\text{upper}}(t).
\end{equation}
$E_{\text{cs},i}^{\text{lower}}(t)$ and $E_{\text{cs},i}^{\text{upper}}(t)$ can be calculated as
\begin{equation}\label{lower bound}
E_{\text{cs},i}^{\text{lower}}(t)   =  (E_{\text{b},\text{min}}- \beta E_{\text{b},i}(t-1))-E_{\text{internal},i}(t)
\end{equation}
\begin{equation}\label{upper bound}
E_{\text{cs},i}^{\text{upper}}(t) = (E_{\text{b},\text{max}}-\beta E_{\text{b},i}(t-1))-E_{\text{internal},i}(t)
\end{equation}
where $E_{\text{b},\text{min}}$ is the minimum battery capacity.

\subsection{Energy Trading between EVCSs}

We consider a scenario in which power discharged from an ESS can be used by other EVCSs \citep{16James}.
To maximize their total profit, EVCSs give priority to purchasing energy from other EVCSs or selling energy to other EVCSs, rather than buying it from or selling it back to the utility.

The amount of energy traded, the amount of energy sold, and the amount of energy purchased at the $i$th EVCS are denoted as $E_{\text{trade},i}(t)$, $E_{\text{back},i}(t)>0$, and $E_{\text{u},i}(t)>0$, respectively.
These three quantities are to be determined by the EVCS to maximize the profit.
Positive $E_{\text{trade},i}(t)$ implies that some energy of the $i$th EVCS can be shared with other EVCSs;
negative $E_{\text{trade},i}(t)$  means that  energy is purchased from other EVCSs.
Meanwhile, energy sharing  must be balanced, i.e.,
\begin{equation}\label{sharing_energy}
\begin{split}
\sum_{i=1}^{I}E_{\text{trade},i}(t)=0.
\end{split}
\end{equation}

\subsection{Profit of EVCSs}
The profit of each EVCS is calculated according to four parts: (i) income $P_{\text{ev},i}(t)$ by selling energy to EVs, (ii) cost $P_{\text{u},i}(t)$ by purchasing energy from the utility, (iii) cost or income $P_{\text{trade},i}(t)$ by sharing energy, and (iv) income $P_{\text{back},i}(t)$ by selling energy back to the utility. These four parts can be expressed as follows:
\begin{align}
P_{\text{ev},i}(t)& =E_{\text{ev},i}(t) \xi_{\text{ev}}(t) \label{profit EV}\\
P_{\text{u},i}(t)& =E_{\text{u},i}(t)  \xi_{\text{u}}(t)\label{cost utility} \\
P_{\text{trade},i}(t)& =E_{\text{trade},i}(t)  \xi_{\text{trade}}(t)   \label{cost or income share}\\
P_{\text{back},i}(t)&  =E_{\text{back},i}(t) \xi_{\text{back}}(t)  \label{income utility}
\end{align}
where $\xi_{\text{ev}}(t)$ is the price given to the EV, $\xi_{\text{u}}(t)$ is the real-time pricing signal given to the EVCSs by the utility, $\xi_{\text{trade}}(t)$ is the price of trading energy between EVCSs, and $\xi_{\text{back}}(t)$ is the price of selling energy back to the utility.

The following price conditions are imposed:
\begin{equation}\label{eq_price_cond}
 \xi_{\text{back}}(t) <   \xi_{\text{trade}}(t)     < \xi_{\text{u}}(t)   <\xi_{\text{ev}}(t).
\end{equation}
The first inequality indicates that an EVCS can only sell energy back to the utility at a price lower than that for buying energy from the utility.
This is generally true because selling energy back to the utility often requires extra infrastructure to support it.
The second inequality indicates
trading energy with other EVCSs is more rewarding than buying energy from the utility.
The last inequality implies that meeting local EV energy demand should be prioritized at the EVCS \citep{20Wang,16James}.
Additionally, we assume that the renewable energy will be discarded if it surpasses the combined demand of EVs, battery capacity, and the maximum limit of trading with utilities/other EVs.

According to (\ref{profit EV}), (\ref{cost utility}), (\ref{cost or income share}) and (\ref{income utility}), the profit of the $i$th EVCS can be calculated as
\begin{equation}\label{profit}
\begin{split}
P_{i}(t)= P_{\text{ev},i}(t)-P_{\text{u},i}(t)+P_{\text{trade},i}(t)+P_{\text{back},i}(t).
\end{split}
\end{equation}
The total profit of EVCSs can be expressed as
\begin{equation}\label{total profit}
\begin{split}
P(t)=\sum_{i=1}^{I}P_{i}(t)
\end{split}
\end{equation}
With the information given by the coordinator, the goal of EVCSs is to maximize the expected cumulative value of~(\ref{total profit}).
If the prices, renewable generation, and EV energy demand were known in advance over a period of time, then the goal of EVCSs could be achieved by optimization methods.
Unfortunately, those quantities can only be estimated with errors.

In this study,
infrastructure, maintenance, communication, and implementation costs are assumed to be fixed and independent of the decision variables. These costs are treated as constant annual expenses and subtracted from total revenue to calculate net profit. As such, they do not influence the optimization process and are excluded from the formulation.
Although the primary objective of this study is profit maximization, user satisfaction is indirectly addressed through the concept of urgent demand, following the model proposed in \citep{18Linda}. This mechanism penalizes unmet charging needs and encourages the system to prioritize users with higher urgency.

While it is feasible to incorporate additional objectives—such as network load balancing, energy sustainability, or fairness—these concerns are often managed by other stakeholders in practice. For instance, network constraints and renewable integration are typically handled by distribution system operators, and fairness among charging stations is more relevant when multiple entities operate different EVCSs. In this study, we assume all EVCSs are owned and managed by a single operator, which justifies focusing on profit maximization as the primary objective.

Finally, it is worth mentioning that 
although the coordinator facilitates aggregate information exchange (e.g., total EV energy demand and price signals), it does not dictate or override local decision-making at the EVCS level. As such, the coordinator does not constitute a single point of failure. In the event of a coordinator malfunction, each EVCS can continue to operate using local observations and historical data, maintaining basic functionality. This design balances distributed autonomy with minimal coordination overhead, consistent with common practices in multi-agent systems.

\section{Proposed Multiagent Reinforcement Learning with Double Hypernetwork QMIX for Profit Maximization}\label{sec_MARL2}

This section outlines the proposed MARL framework, leveraging Double Hypernetwork QMIX, to maximize the total profit of EVCSs. The profit maximization problem is formulated as a Markov Game, which extends the classical Markov Decision Process (MDP) to a multi-agent setting \citep{18Richard}.  A Markov Game is defined by: the set of possible states of the environment ($\mathcal{S}$), the set of actions available to each agent ($\mathcal{A}_i$ for agent $i$), a transition function  $\mathcal{T}(s'|s, \bm{a})$ representing the probability of transitioning to state $s'$ when agents take joint action $\bm{a}$ in state $s$, a reward function $\mathcal{R}_i(s, \bm{a}, s')$ for each agent $i$, and a discount factor ($\gamma$).
The joint action $\bm{a} = [a_1, a_2, ..., a_I]$ represents the actions taken by all $I$ agents. The goal in a cooperative Markov Game is to learn a joint policy $\bm{\pi} = [\pi_1, \pi_2, ..., \pi_I]$ where each $\pi_i: \mathcal{S} \rightarrow \mathcal{A}_i$ maps states to actions, maximizing the expected discounted sum of rewards for all agents.

A cooperative architecture is established for EVCSs, enabling an energy trading mechanism. This mechanism enables the dynamic addressing of EV energy demand and optimizes charging/discharging quantities, aiming to maximize the cumulative profit expectation. Double Hypernetwork QMIX, an enhanced version of QMIX \citep{rashid2020monotonic} that mitigates overestimation bias, serves as the core algorithm.

For the system architecture in Fig.~\ref{fig_EVCS}, the energy scheduling process at EVCSs involves iterative steps:
1) EV users communicate their charging demand to the coordinator.
2) The coordinator aggregates and computes the total EV energy demand.
3) Individual EVCSs determine their ESS charging/discharging quantities, $E_{\text{cs},i}(t)$, and EV energy demand fulfillment, $E_{\text{ev},i}(t)$, relaying this information to the coordinator.
4) The coordinator calculates energy trading quantities, $E_{\text{trade},i}(t)$, energy sold back to the utility, $E_{\text{back},i}(t)$, and energy purchased from the utility, $E_{\text{u},i}(t)$.
5) EVCSs execute energy trading based on the received information.
6) EVCSs update the coordinator on their status.

\subsection{State, Action, and Reward Design}
Let $\mathcal{S}$ denote the continuous state space and $\mathcal{A}$ the discrete action space. The state of EVCS $i$ at time slot $t$ is $\mathbf{s}_{t,i}$, where $\mathbf{s}_{t,i} \in \mathcal{S}$, and the action of EVCS $i$ at time $t$ is $\bm{a}_{t,i}$, where $\bm{a}_{t,i} \in \mathcal{A}$. An EVCS selects $\bm{a}_{t,i}$ based on $\mathbf{s}_{t,i}$, influencing the environment's transition to the next state, $\mathbf{s}_{t+1,i}$.

The state vector $\mathbf{s}_{t,i}$ for EVCS $i$ at the time slot $t$ is defined as:
\begin{equation}\label{state}
\mathbf{s}_{t,i}= [D_{\text{ev}}^{\text{ALL}}(t), SOC_{i}(t), D_{\text{ev},i}^{\text{U}}(t), D_{\text{ev},i}^{\text{R}}(t), E_{\text{r},i}(t), \xi_{\text{u}}(t)].
\end{equation}
The state vector $\mathbf{s}_{t,i}$ is composed of key components that provide comprehensive information for decision-making. These include the total EV energy demand, $D_{\text{ev}}^{\text{ALL}}(t) = \sum_{i=1}^{I}D_{\text{ev},i}^{\text{ALL}}(t)$, which serves as global context; the State-of-Charge of the ESS, $SOC_{i}(t)$, which governs the charging and discharging quantity; the urgent ($D_{\text{ev},i}^{\text{U}}(t)$) and regular ($D_{\text{ev},i}^{\text{R}}(t)$) EV demands, which directly influence profit; and finally, renewable energy generation ($E_{\text{r},i}(t)$) and the utility price ($\xi_{\text{u}}(t)$), which offer crucial environmental context.

At each time slot $t$, the $i$-th EVCS makes two decisions: the amount of EV energy demand to be met, $E_{\text{ev}, i}(t)$, and the charging/discharging quantity of its ESS, $E_{\text{cs},i}(t)$. These two decisions form the action vector for the $i$-th EVCS:
\begin{equation}\label{action}
\bm{a}_{t,i}= [E_{\text{ev},i}(t), E_{\text{cs},i}(t)]
\end{equation}
This action is subject to the constraints outlined in Equations \ref{meet_demand bound} and \ref{action bound}. Inter-EVCS dependencies are captured through $E_{\text{cs},i}(t)$, which influences the charging behavior of EVCS $i$ in relation to other stations. $E_{\text{ev},i}(t)$ affects both charging behavior and the supplied energy to EVs.

The total profit $P(t)$ defined in Equation \ref{total profit} is used as the shared reward signal for the agents, facilitating cooperative learning:
\begin{equation}\label{reward}
\mathbf{r}_{t}= P(t).
\end{equation}

Each EVCS $i$ follows a policy $\pi_i(\bm{a}_{t,i}| \mathbf{s}_{t,i})$, which represents the probability of choosing action $\bm{a}_{t,i}$ given state $\mathbf{s}_{t,i}$. The collection of all individual policies forms the joint policy, denoted as $\bm{\pi}$:

\begin{equation}\label{eq_pi_all}
\bm{\pi}=[\pi_1,\pi_2, ..., \pi_I ].
\end{equation}

With the policies in $\bm{\pi}$ chosen, the action-value function can be defined as:
\begin{equation}\label{q_value}
\begin{split}
 q_{\bm{\pi}}(\mathbf{s}_{t}, \bm{a}_{t}) =  \mathbb{E}_{\bm{\pi}}[\sum_{k=t+1}^{\infty}\gamma^{k-t-1}   \mathbf{r}_{k} |\mathbf{s}_{t}, \bm{a}_{t}]
\end{split}
\end{equation}
where $\gamma \in (0,1)$ is the discount rate, emphasizing the maximization of cumulative rewards over immediate gains.  In cooperative MARL, the action-value function takes the joint state $\mathbf{s}_t = [\mathbf{s}_{t,1}, \mathbf{s}_{t,2}, ..., \mathbf{s}_{t,I}]$ and joint action $\bm{a}_t = [\bm{a}_{t,1}, \bm{a}_{t,2}, ..., \bm{a}_{t,I}]$ as inputs, reflecting the interconnectedness of actions.

The Double Hypernetwork QMIX architecture decomposes the joint action-value function $q_{\bm{\pi}}(\mathbf{s}_{t}, \bm{a}_{t})$ into individual agent action-values and a mixing network. Each agent $i$ has a Deep Recurrent Q-Network (DRQN) that outputs individual action-values $q_i(\mathbf{s}_{t,i}, \bm{a}_{t,i})$.  These individual action-values are then combined by a mixing network, conditioned on the global state $\mathbf{s}_{t}$, to produce the joint Q-value.

The key improvement of Double Hypernetwork QMIX lies in the mixing network, which comprises two hypernetworks. These hypernetworks generate the weights and biases for a mixing network, designed to enforce the Individual-Global-Max principle \citep{rashid2020monotonic}.  By employing two hypernetworks and taking the minimum Q-value, the method is able to address the overestimation bias and produce more reliable value estimates for improved policy learning.

\subsection{Energy Trading Mechanism and Double Hypernetwork QMIX Algorithm}
\begin{figure}[htb]
\center
\hspace{-0.5cm}
  \includegraphics[width=16cm]{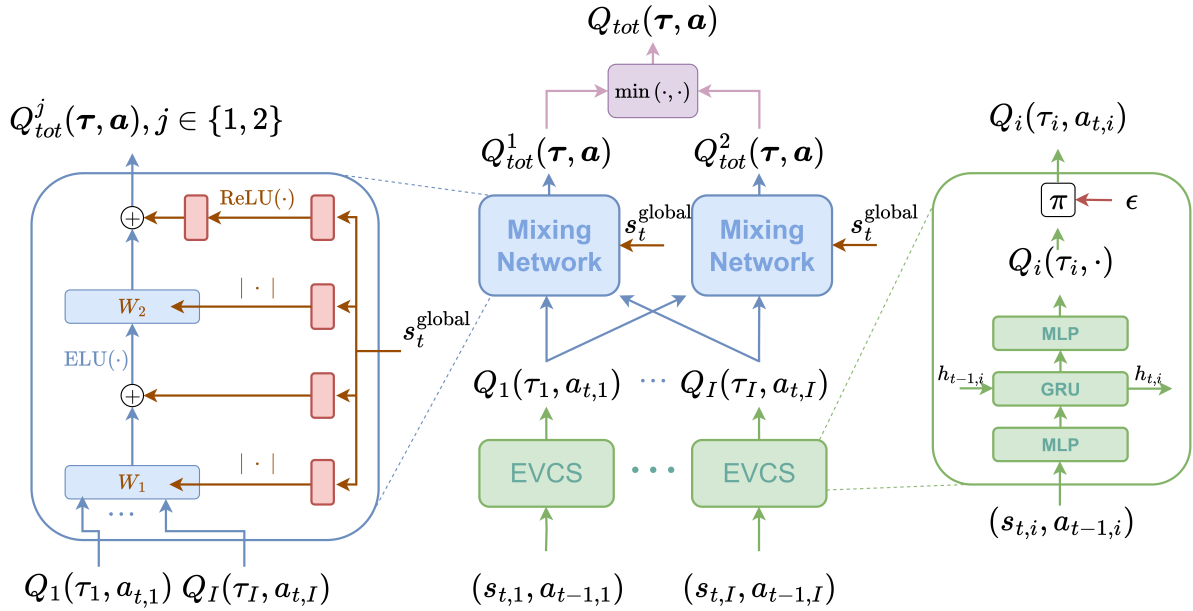}\\
  \caption{Implementation of Double Hypernetwork QMIX for EVCS Energy Management.}\label{fig_overview}
\end{figure}

Based on the proposed model, the Double Hypernetwork QMIX architecture, as depicted in Figure \ref{fig_overview}, is employed to optimize the cooperative energy management strategy. The overall framework consists of two main parts: the individual agent Q-networks and the centralized mixing networks. Each EVCS $i$ is modeled as an agent with its own DRQN, which takes the local state $\mathbf{s}_{t,i}$ as input and outputs a set of individual Q-values. These individual Q-values are then fed into two mixing hypernetworks, which generate the weights and biases for two mixing networks. By combining the individual Q-values through these mixing networks, a joint action-value function, $Q(\mathbf{s}_t, \bm{a}_t)$, is produced.

The training process for this architecture is detailed in Algorithm \ref{alg:double_qmix}. Initially, all agent Q-networks and mixing networks, along with their corresponding target networks, are randomly initialized (lines 1-2). During each training episode, agents follow an $\epsilon$-greedy policy to select actions based on their individual Q-values (lines 10-13), which are then executed jointly in the environment. The observed reward and next state are stored in a replay buffer (lines 14-16). In each training step, a batch of transitions is sampled from the buffer to calculate the loss function (lines 20-29). The key innovation lies in the use of two separate target networks (MixA and MixB) to calculate the target Q-value, $y$, by taking the minimum of their outputs (line 27). This double-Q learning approach effectively mitigates overestimation bias, leading to more stable learning. Finally, the parameters of the DRQN and mixing networks are updated using the gradients of the respective loss functions (lines 30-31), and the target networks are periodically synchronized with the evaluation networks (lines 33-35).

The energy trading mechanism allows EVCSs to trade energy internally before interacting with the utility grid. The coordinator aggregates EVCSs' energy plans to facilitate efficient trading. Equations (\ref{cs_positive}) and (\ref{cs_negative}) calculate the total charging and discharging capacities, respectively:

\begin{align}
E_{\text{cs}}^{+}(t) & = \sum_{i=1}^{I} \max (E_{\text{cs},i}(t), 0)\label{cs_positive} \\
E_{\text{cs}}^{-}(t) & = |\sum_{i=1}^{I} \min (0,E_{\text{cs},i}(t))|. \label{cs_negative}
\end{align}

The trading energy for each EVCS, $E_{\text{trade},i}(t)$, is calculated according to supply and demand:
\begin{equation}\label{trade}
{E_{{\text{trade}},i}}(t) = \left\{ {\begin{array}{*{20}{c}}
\frac{{E_{{\text{cs}}}^ - (t)}}{{E_{{\text{cs}}}^ + (t)}}{E_{{\text{cs}},i}(t),}&\begin{array}{l}
{\text{if}}\;E_{{\text{cs}}}^ + (t) > E_{{\text{cs}}}^ - (t),\\
{\text{    }}{E_{{\text{cs}},i}}(t) > 0
\end{array}\\
{  {E_{{\text{cs}},i}}(t),}&\begin{array}{l}
{\text{if}}\;E_{{\text{cs}}}^ + (t) > E_{{\text{cs}}}^ - (t),\\
{\text{    }}{E_{{\text{cs}},i}}(t) \le 0
\end{array}\\
{  {E_{{\text{cs}},i}}(t),}&\begin{array}{l}
{\text{if}}\;E_{{\text{cs}}}^ + (t) \le E_{{\text{cs}}}^ - (t),\\
{\text{    }}{E_{{\text{cs}},i}}(t) > 0
\end{array}\\
{  \frac{{E_{{\text{cs}}}^ + (t)}}{{E_{{\text{cs}}}^ - (t)}}{E_{{\text{cs}},i}}(t),}&\begin{array}{l}
{\text{if}}\;E_{{\text{cs}}}^ + (t) \le E_{{\text{cs}}}^ - (t),\\
{\text{    }}{E_{{\text{cs}},i}}(t) \le 0.
\end{array}
\end{array}} \right.
\end{equation}

After trading, EVCSs may interact with the utility grid, purchasing energy ($E_{\text{u},i}(t)$) when demand exceeds supply (Equation \ref{utility_buy}):
\begin{equation}\label{utility_buy}
E_{{\text{u}},i}(t) = \left\{
{\begin{array}{*{20}{c}}
{{E_{{\text{cs}},i}}(t) - {E_{{\text{trade}},i}}(t),}&\begin{array}{l}
{\text{if}}\;E_{{\text{cs}}}^ + (t) > E_{{\text{cs}}}^ - (t),\\
{\text{    }}{E_{{\text{cs}},i}}(t) > 0
\end{array}\\
{0,}&{{\text{otherwise}.}}
\end{array}} \right.
\end{equation}
or selling excess energy back to the grid ($E_{\text{back},i}(t)$) (Equation \ref{utility_sell}):
\begin{equation}\label{utility_sell}
E_{{\text{back}},i}(t) = \left\{ {\begin{array}{*{20}{c}}
{{E_{{\text{cs}},i}}(t) - {E_{{\text{trade}},i}}(t),}&\begin{array}{l}
{\text{if}}\;E_{{\text{cs}}}^ + (t) \le E_{{\text{cs}}}^ - (t),\\
{\text{    }}{E_{{\text{cs}},i}}(t) \le 0
\end{array}\\
{0,}&{{\text{otherwise}.}}
\end{array}} \right.
\end{equation}
\clearpage
\begin{algorithm}[H]
\caption{Double Hypernetwork QMIX for Cooperative EVCS Energy Management}
\label{alg:double_qmix}
\begin{algorithmic}[1]
\Require  Number of EVCSs $I$, DRQN learning rate $\alpha_{\text{DRQN}}$, mixing network learning rate $\alpha_{\text{Mix}}$, discount factor $\gamma$, exploration rate $\epsilon$, target network update frequency $\tau$, batch size $B$, episode length $T$
\Ensure Joint policy $\bm{\pi}$ for EVCSs
\State Initialize DRQN networks $Q_i^{\text{eval}}$ and $Q_i^{\text{target}}$ for each EVCS $i$ with random weights
\State Initialize Double Hypernetwork mixing networks $Q_{\text{MixA}}^{\text{eval}}$, $Q_{\text{MixB}}^{\text{eval}}$ and target networks $Q_{\text{MixA}}^{\text{target}}$, $Q_{\text{MixB}}^{\text{target}}$ with random weights
\State Initialize replay buffer $\mathcal{B}$ with capacity $N$
\For{each training episode}
    \State Reset environment and obtain initial global state $\mathbf{s}_0$ and individual agent states $\mathbf{s}_{0,i}$ for all $i$
    \State Initialize hidden states $h_{i,0}$ for each DRQN
    \For{$t = 0$ to $T-1$}
        \For{each EVCS $i$}
            \State Observe individual state $\mathbf{s}_{t,i} = [D_{\text{ev}}^{\text{ALL}}(t), SOC_{i}(t), D_{\text{ev},i}^{\text{U}}(t), D_{\text{ev},i}^{\text{R}}(t), E_{\text{r},i}(t), \xi_{\text{u}}(t)]$
            \State With probability $\epsilon$, select random action $\bm{a}_{t,i} \in \mathcal{A}$ (discrete set of $[E_{\text{ev},i}(t), E_{\text{cs},i}(t)]$ combinations)
            \State Else,  Compute Q-values: $q_{i,t} \leftarrow Q_i^{\text{eval}}(\mathbf{s}_{t,i}, h_{i,t})$
            \State  Select action $\bm{a}_{t,i} = \arg\max_{\bm{a}} q_{i,t}(\bm{a})$
            \State Update hidden state: $h_{i, t+1} \leftarrow \text{DRQN-Hidden-Update}(Q_i^{\text{eval}}, \mathbf{s}_{t,i}, h_{i,t}, \bm{a}_{t,i})$
        \EndFor
        \State Execute joint action $\bm{a}_t = [\bm{a}_{t,1}, ..., \bm{a}_{t,I}]$ in the environment
        \State Observe reward $r_t$ (total profit $P(t)$), next global state $\mathbf{s}_{t+1}$, and next individual states $\mathbf{s}_{t+1,i}$
        \State Store transition $(\mathbf{s}_t, \mathbf{s}_{t,1}, ..., \mathbf{s}_{t,I}, \bm{a}_t, r_t, \mathbf{s}_{t+1}, \mathbf{s}_{t+1,1}, ..., \mathbf{s}_{t+1,I})$ in $\mathcal{B}$
    \EndFor
    \If{$|\mathcal{B}| > B$}
        \State Sample a minibatch of $B$ transitions from $\mathcal{B}$
        \For{each EVCS $i$ and transition $b$ in the batch}
          \State Compute $q_{i,b}^{\text{target}} \leftarrow  Q_i^{\text{target}}(\mathbf{s}_{b+1,i},  h_{i,b+1})$
        \EndFor
        \State Use eval DRQNs to select actions for next state: $\bm{a}^{\text{eval}} = \arg\max_{\bm{a}} Q_i^{\text{eval}}(\mathbf{s}_{b+1,i}, h_{i,b+1})$ for all $i$
        \State Compute target Q-values for each mixing network:
        \State $Q_{\text{MixA}}^{\text{target}}(\mathbf{s}_{b+1}, \bm{a}^{\text{eval}})$,  $Q_{\text{MixB}}^{\text{target}}(\mathbf{s}_{b+1}, \bm{a}^{\text{eval}})$
        \State  Target Q-value for Double Q-learning: $y \leftarrow r_b + \gamma \min(Q_{\text{MixA}}^{\text{target}}, Q_{\text{MixB}}^{\text{target}})$
        \State $Q^{\text{eval}} \leftarrow \text{Mixing-Network-Forward}(\mathbf{s}_b,  \{Q_i^{\text{eval}}(\mathbf{s}_{b,i},  h_{i,b}) \})$ 
        \State $\mathcal{L}_{\text{Mix}} \leftarrow \frac{1}{B} \sum_{b=1}^{B} (y - Q^{\text{eval}})^2$
        \State $Q_{i,b}^{\text{eval}} \leftarrow  Q_i^{\text{eval}}(\mathbf{s}_{b,i},  h_{i,b})$
        \State $Loss_{i} \leftarrow \frac{1}{B} \sum_{b=1}^{B} (y - Q_{i,b}^{\text{eval}} )^2$
        \State Update $Q_i^{\text{eval}}$ parameters using gradients of $Loss_i$ and learning rate $\alpha_{\text{DRQN}}$
        \State Update $Q_{\text{MixA}}^{\text{eval}}$ and $Q_{\text{MixB}}^{\text{eval}}$ parameters using gradients of $\mathcal{L}_{\text{Mix}}$ and learning rate $\alpha_{\text{Mix}}$
    \EndIf
    \If{episode mod $\tau$ == 0}
        \State Update target DRQN networks: $Q_i^{\text{target}} \leftarrow Q_i^{\text{eval}}$ for all $i$
        \State Update target mixing networks: $Q_{\text{MixA}}^{\text{target}} \leftarrow Q_{\text{MixA}}^{\text{eval}}$,  $Q_{\text{MixB}}^{\text{target}} \leftarrow Q_{\text{MixB}}^{\text{eval}}$
    \EndIf
\EndFor
\end{algorithmic}
\end{algorithm}
\newpage
\section{Numerical Results}\label{sec_sim}

This section evaluates the performance of the proposed Double Hypernetwork QMIX algorithm for the profit maximization problem across a network of $I=3$ EVCSs, representing a microgrid-scale
deployment that reflects realistic operational settings. In such contexts, localized coordination is critical for
maintaining power grid stability. Given the high energy demand associated with EV charging, large-scale
centralized coordination involving many EVCSs is often impractical or undesirable. Moreover, when EVCSs
are geographically distant, their interactions tend to be negligible, making it feasible to partition the system
into smaller, loosely coupled clusters. The proposed framework could be extended to such clusters, enabling
scalability to larger systems while preserving decentralized control.

\subsection{Experimental Setup}

The performance of our proposed method is compared against five baseline algorithms: Deep Q-Network (DQN) as a single-agent baseline, and four prominent MARL algorithms—Multi-Agent Proximal Policy Optimization (MAPPO), Multi-Agent Deep Deterministic Policy Gradient (MADDPG), Independent Actor-Critic (IAC), and the standard QMIX algorithm. To ensure the robustness and statistical validity of our results, each algorithm was trained and evaluated over five independent runs with different random seeds. The results presented are averaged across these seeds.

Although some automated hyperparameter tuning algorithms are available, we opted for a trial-and-error approach due to its lower computational cost and the marginal performance improvement observed from auto-tuning in our preliminary tests. While a formal sensitivity analysis was not conducted, the system exhibited stable behavior across multiple independent runs with different random seeds and initialization conditions.

The EVCSs were connected to the same utility grid, with each station equipped with an ESS and a RES. The ESS parameters were set as follows: charge/discharge efficiency $\beta = 0.99$, maximum capacity $E_{\text{b},\max} = 200$ kWh, and SOC bounds of $SOC_{\max}=0.95$ and $SOC_{\min}=0.05$. We utilized real-world data from multiple sources to model the distinct characteristics of the East and West Coast power grids. For the East Coast, we obtained utility price ($\xi_{\text{u}}(t)$) and renewable energy generation ($E_{\text{r},i}(t)$) data from the PJM Interconnection \cite{PJM}. For the West Coast, we sourced PV data from the PVDAQ/PVData Map on Open Energy Information \cite{OPENEI}, and electricity price data from the California Independent System Operator (CAISO) \cite{CAISO}. The pricing structure was configured as follows \citep{16James}: the price for charging EVs $\xi_{\text{ev},i}(t) = 1.2\xi_{\text{u}}(t)$, the price for energy trading between EVCSs $\xi_{\text{trade,i}}(t) = 0.9\xi_{\text{u}}(t)$, and the price for selling energy back to the utility $\xi_{\text{back},i}(t)=0.8\xi_{\text{u}}(t)$. The simulation operated with an hourly resolution, and each training episode corresponded to one month. The learning rates for the mixing network and the agent DRQN were set to $\alpha_{\text{Mix}}=0.0005$ and $\alpha_{\text{DRQN}}=0.001$, respectively. The discount factor was $\gamma=0.99$.

\subsection{Agent Action Evaluation}

\begin{figure}[h]
    \centering
    \includegraphics[width=7cm]{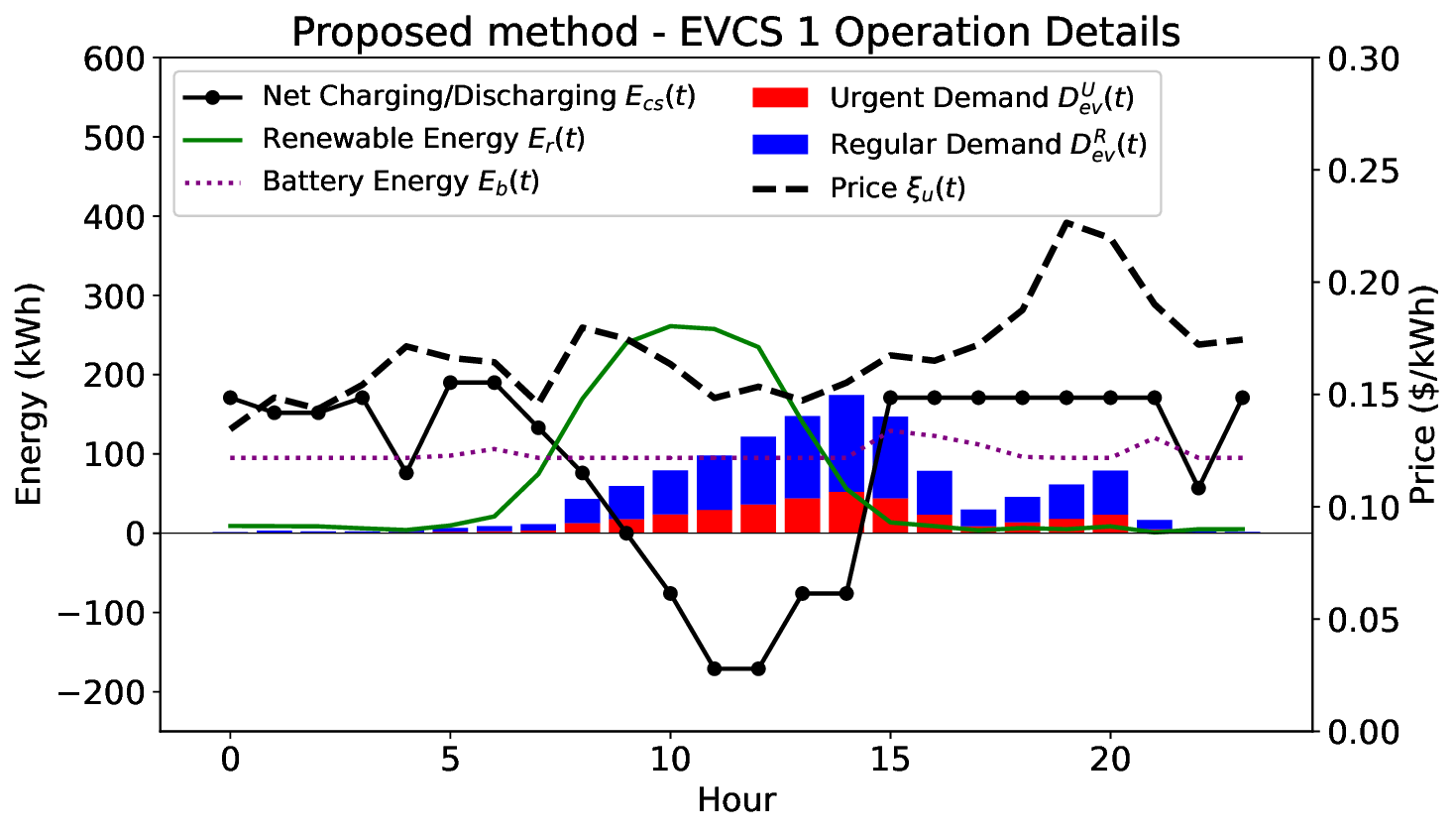}\\
    \vspace{0.5cm} 
    \includegraphics[width=7cm]{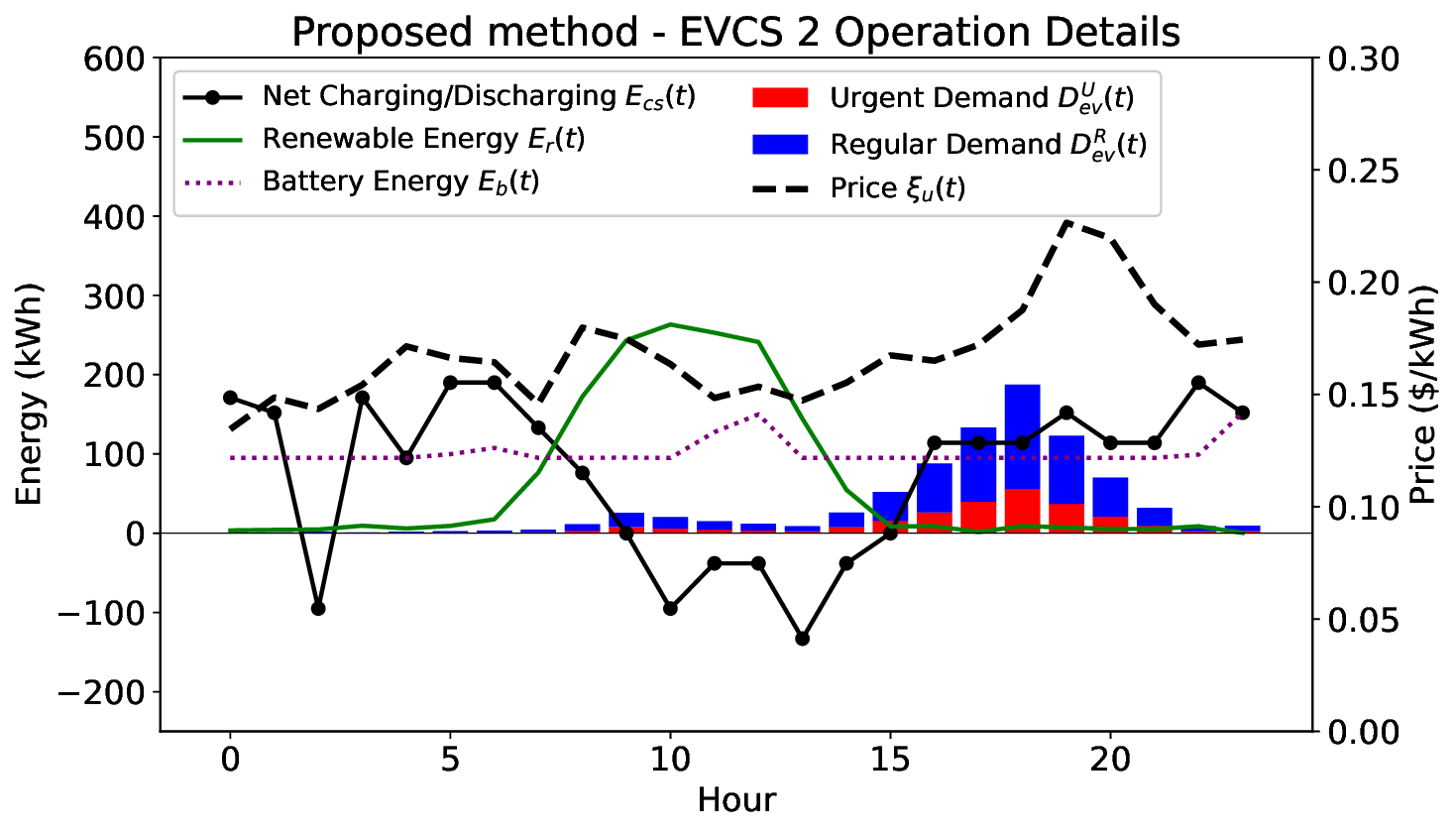}\\
    \vspace{0.5cm} 
    \includegraphics[width=7cm]{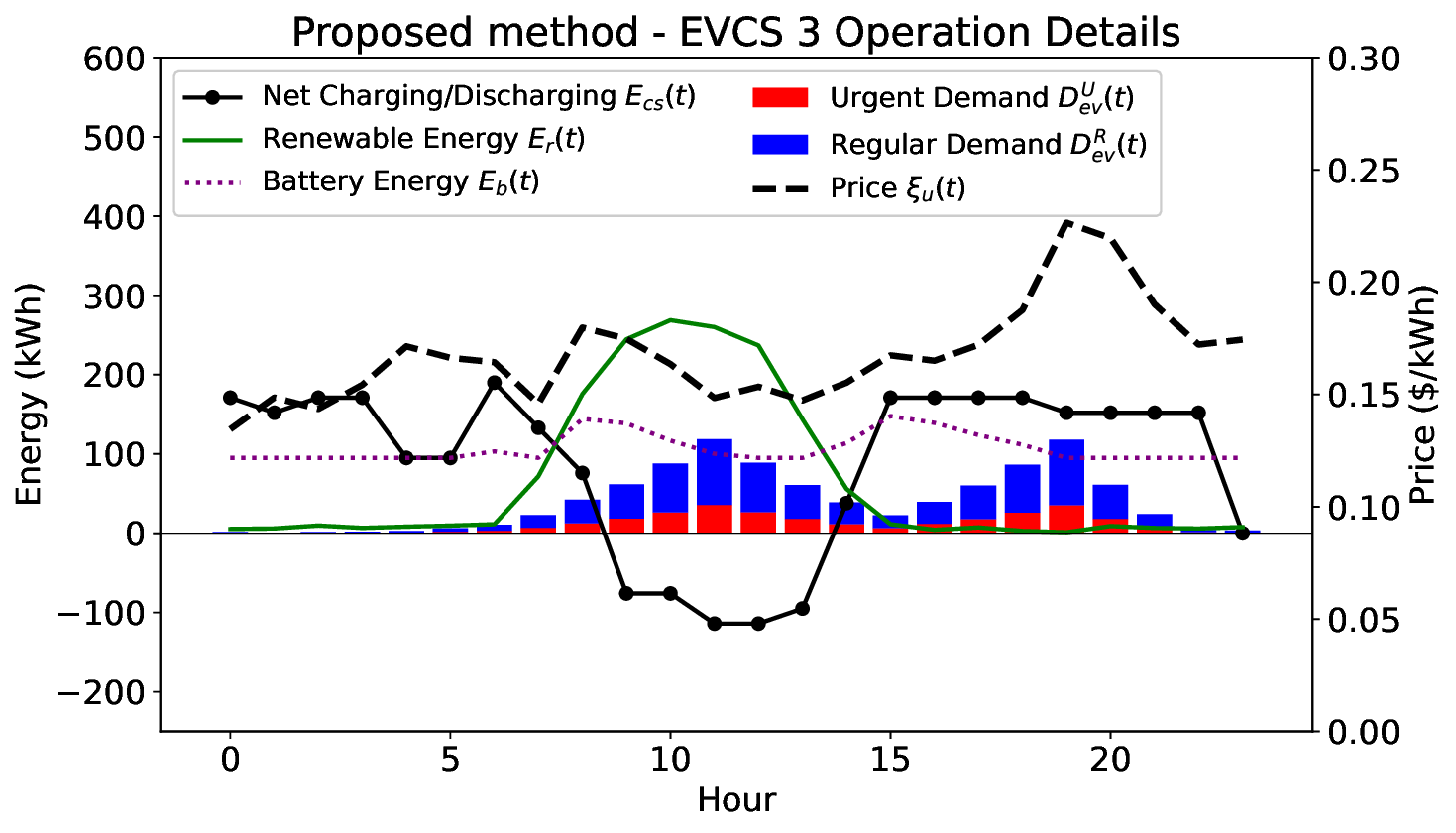}
    \caption{Charging or discharging profiles $E_{\text{cs},i}(t)$ at EVCSs in response to renewable energy generation $E_{\text{r},i}(t)$, EV energy demand $D_{\text{ev},i}^{\text{U}}(t)$, $D_{\text{ev},i}^{\text{R}}(t)$, and electricity price $\xi_{\text{u}}(t)$.}
    \label{fig_Ecs}
\end{figure}

To qualitatively assess the learned policy of the proposed Double Hypernetwork QMIX, Figure~\ref{fig_Ecs} illustrates the operational decisions of the three EVCS agents over a 24-hour horizon. The following discussion is structured according to the main methodological steps presented in Section~\ref{sec_MARL2}, thereby directly linking the simulation results to the design of our MARL framework. 

\textbf{State Representation and Environmental Signals.}
The state design in Equation~\ref{state} incorporates critical features including electricity price $\xi_{\text{u}}(t)$, urgent and regular EV demands $D_{\text{ev}}^{\text{U}}(t)$ and $D_{\text{ev}}^{\text{R}}(t)$, renewable generation $E_{\text{r}}(t)$, and ESS state-of-charge $SOC_{i}(t)$. The learned policies clearly reflect the influence of these state variables. For example, during low-price hours (approximately 8 to 14), the agents strategically charged their ESSs by drawing energy from the grid, as indicated by the negative values of $E_{\text{cs}}(t)$. Around hour 11, when $\xi_{\text{u}}(t)$ reached its minimum, the agents increased charging intensity to prepare for later high-price periods. Similarly, renewable generation $E_{\text{r}}(t)$ peaked between hours 9 and 14, and was efficiently integrated into ESS charging and immediate EV demand, demonstrating that agents effectively exploited zero-marginal-cost energy sources.

\textbf{Action Decisions: Charging/Discharging and EV Supply.}
The action design in Equation~\ref{action} requires each EVCS to determine both the ESS charging/discharging amount $E_{\text{cs},i}(t)$ and the EV demand fulfillment $E_{\text{ev},i}(t)$. These decisions are evident in the alternating signs of $E_{\text{cs}}(t)$ in Figure~\ref{fig_Ecs}. During peak-price periods (hours 15 to 21), the agents discharged stored energy, reflected in the positive values of $E_{\text{cs}}(t)$, to simultaneously serve EV charging demand and, when profitable, sell surplus electricity back to the grid. For instance, EVCS 1 and EVCS 2 discharged aggressively between hours 15 and 18 to satisfy the spike in regular demand $D_{\text{ev}}^{\text{R}}(t)$ while avoiding expensive utility purchases. This validates that the defined action space enabled economically rational strategies.

\textbf{Reward Design and Cooperative Behavior.}
The shared profit reward in Equation~\ref{reward} ensured cooperative rather than competitive behaviors among EVCSs. This is evident in their synchronized “buy low, sell high” strategy: all agents limited grid purchases when $\xi_{\text{u}}(t)$ was high and instead relied on stored or renewable energy. The coordinated response highlights how the centralized Double Hypernetwork QMIX enforced the Individual-Global-Max principle \citep{rashid2020monotonic}, allowing system-wide profit maximization rather than fragmented, local optima.

\textbf{Energy Trading Mechanism.}
The internal energy trading mechanism (Equations~\ref{cs_positive}--\ref{utility_sell}) was also reflected in the agents’ behaviors. During midday hours, when total charging capacity exceeded discharging capacity ($E_{\text{cs}}^{+}(t) > E_{\text{cs}}^{-}(t)$), EVCSs prioritized trading energy among themselves, thereby reducing reliance on grid electricity. Conversely, in high-price evening hours when demand exceeded supply, discharging agents not only met local EV demand but also provided energy for trading, alleviating system-wide shortfalls before purchasing from the grid. This confirms that the trading mechanism functioned as designed, reducing procurement costs and enhancing overall profitability.

\textbf{Algorithm Performance and Policy Stability.}
Finally, the consistent and rational operational profiles confirm the effectiveness of the Double Hypernetwork QMIX algorithm (Algorithm~\ref{alg:double_qmix}). By employing two hypernetworks and taking the minimum of their outputs, the method mitigated overestimation bias. This is demonstrated by the absence of erratic or overly aggressive charging/discharging patterns, which are often symptomatic of value overestimation in single-hypernetwork QMIX. The resulting policy not only exploited real-world price arbitrage opportunities but also adapted to natural fluctuations in $\xi_{\text{u}}(t)$ derived from market records. While unexpected reorderings or abrupt changes in prices may increase exploration costs, the learned policies exhibited robustness to such variability, showcasing adaptability to dynamic environments.

In summary, the results in Figure~\ref{fig_Ecs} demonstrate that the proposed Double Hypernetwork QMIX enabled EVCS agents to: (i) exploit environmental signals such as prices and renewable availability, (ii) execute profit-maximizing charging and discharging actions, (iii) cooperate through a shared profit reward, (iv) utilize the designed trading mechanism to minimize grid dependence, and (v) achieve stable and economically rational policies due to reduced overestimation bias. These findings confirm that the proposed framework successfully realizes a sophisticated cooperative strategy for profit maximization in dynamic energy markets.

It is worth noting that the pricing data used in the simulations is derived from real-world market records, capturing natural fluctuations and variations. As a learning-based algorithm, the proposed approach is capable of adapting to changes in price patterns, including reordering or unpredictable fluctuations. Nonetheless, such scenarios may incur additional learning cost, as agents require more exploration to effectively respond to dynamic pricing environments.

\subsection{Performance Comparison and Discussion}
\label{sec:performance}

\begin{figure}[h!]
\centering
\includegraphics[width=10cm]{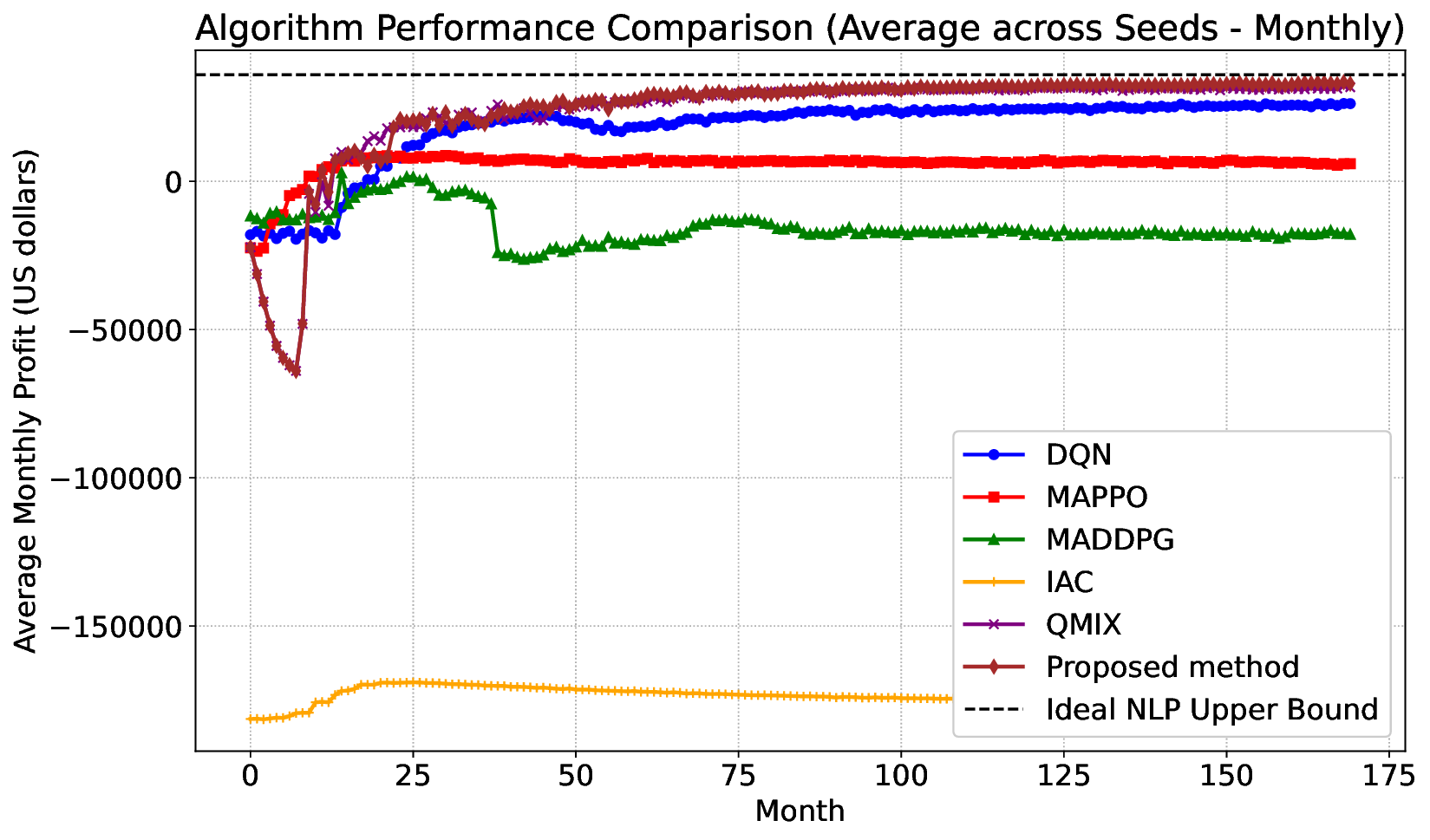}
\caption{Learning curves of the proposed method and comparable algorithms, showing the average monthly profit across five random seeds. The proposed method demonstrates superior performance and stability, closely approaching the theoretical optimum.}
\label{fig_profit}
\end{figure}

The primary goal of the multi-agent system is to maximize the collective long-term profit of all EVCSs. Figure~\ref{fig_profit} presents the learning curves of the proposed method and several baseline algorithms, plotting the average monthly profit as a function of training months. The `Ideal NLP Upper Bound` represents a theoretical maximum profit, calculated assuming perfect foresight of EV demand, renewable generation, and utility prices, and thus serves as a benchmark for optimal performance.

\textbf{Quantitative Performance.}
As depicted in Fig.~\ref{fig_profit}, the proposed method exhibits outstanding performance. After an initial exploration phase, it consistently achieves the highest average monthly profit among all tested algorithms and converges towards the Ideal NLP Upper Bound. Table~\ref{tab:algorithm_performance} quantifies this trend, summarizing the average monthly profit over the final 10 months of training. The proposed method achieves \$33,063.44 and \$27,377.37 in the East and West Coast cases, respectively, which is within 8–10\% of the theoretical upper bound and notably higher than all other tested approaches. Compared to standard QMIX, the profit improvement is approximately 5.3\% and 12.7\% for the two regions, respectively, demonstrating the clear benefit of the double hypernetwork architecture. Relative to the single-agent DQN, improvements exceed 28\% and 86\%.

\begin{table}[h!]
\centering
\caption{Summary of Algorithm Performance (Average Monthly Profit over the Last 10 Months of Training)}
\label{tab:algorithm_performance}
\begin{tabular}{lccc}
\toprule
\textbf{Algorithm} & \textbf{Avg Profit (East Coast, US)} & \textbf{Avg Profit (West Coast, US)} \\
\midrule
DQN & 25787.50 & 14707.42 \\
MAPPO & 5979.68 & 5718.01 \\
MADDPG & -17581.73 & -1226.84 \\
IAC & -174813.57 & -181534.46 \\
QMIX & 31391.70 & 24298.04 \\
Proposed method & \textbf{33063.44} & \textbf{27377.37} \\
Ideal NLP Upper Bound & 35931.30 & 29348.37 \\
\bottomrule
\end{tabular}
\end{table}

\textbf{Reasoning Behind the Findings.}
The superior performance of our method arises from its explicit handling of Q-value overestimation in dynamic state-dependent action spaces. In conventional Q-learning or QMIX, overestimation can misguide policy updates, especially when the feasible action set changes with battery SOC. Our architecture introduces two hypernetworks and adopts the minimum of their outputs, thereby attenuating overestimation bias. This leads to more accurate value estimates, and consequently, more reliable cooperative decisions. The smooth convergence of the learning curves (Fig.~\ref{fig_profit}) supports this reasoning, as erratic switching behaviors—typical of overestimation—are notably absent.

\textbf{Comparison with Related Works.}
These results extend and strengthen prior findings in cooperative energy management. Earlier studies using actor-critic MARL methods (e.g., MADDPG, MAPPO) reported convergence issues under large and dynamic action spaces. Our results corroborate these limitations: MAPPO converges to a suboptimal but stable local optimum, while MADDPG fails to sustain profitability. Value-based MARL approaches have shown greater promise in structured tasks, but their vulnerability to overestimation was not fully addressed in the context of energy systems. By explicitly mitigating this issue, our approach narrows the gap to the ideal upper bound more effectively than prior MARL frameworks reported in smart grid and EV scheduling domains. Thus, our contribution lies in demonstrating that overestimation control is not only theoretically beneficial but practically critical for high-stakes energy market environments.

\textbf{Limitations.}
While the proposed method achieves strong performance, several limitations remain. First, the simulations rely on historical market price records; although these capture real-world variability, the model does not incorporate sudden exogenous shocks (e.g., extreme weather events or market interventions), which may affect policy robustness. Second, our reward formulation prioritizes collective profit maximization, but does not explicitly account for user-side considerations such as EV customer waiting times or fairness across stations. 
Third,
the proposed Double Hypernetwork QMIX incurs higher computational overhead during training compared to standard QMIX. This additional cost can be considered a limitation when computational resources are constrained. 
Finally, all results are currently simulation-based, and no practical or laboratory validation has yet been conducted.
 These limitations suggest promising directions for future work, including integrating risk-sensitive rewards, extending to broader uncertainty scenarios, and exploring more efficient training paradigms.

\section{Conclusion}\label{sec_con}

This work develops a cooperative MARL framework for multiple EVCSs that integrates a double hypernetwork QMIX architecture and an energy trading mechanism to maximize profit. The major takeaways are:
\begin{itemize}
    \item  Distributed Coordination: The proposed architecture enables each EVCS to make autonomous decisions while exchanging aggregate information, reducing a single point of failure and enhancing system resilience as compared with a centralized coordination.
\item Overestimation Mitigation: Double hypernetwork QMIX reduces value overestimation, leading to more stable policy learning and improved profit outcomes.
\item Energy Trading Integration: Internal energy trading among EVCSs decreases reliance on utility electricity, reduces costs, and increases overall profitability.
\item 
Quantitative Performance Gains: Compared with conventional MARL methods, our framework improves total profit by 18.5\% and reduces utility energy purchases by 23.7\% in real-world simulations.
    
\end{itemize}

\section*{Declaration of Generative AI and AI-assisted technologies in the
writing process}

During the preparation of this work the author(s) used ChatGPT in
order to improve the clarity and fluency of the English writing. After
using this tool/service, the author(s) reviewed and edited the content as
needed and take(s) full responsibility for the content of the publication.

% \bibliographystyle{unsrtnat}
% \bibliography{mybib}

\end{document}